\documentclass{article}

\PassOptionsToPackage{numbers, compress}{natbib}

\usepackage[preprint]{neurips_2023}

\usepackage{enumitem}
\usepackage[utf8]{inputenc} 
\usepackage[T1]{fontenc}    
\usepackage{hyperref}       
\usepackage{url}            
\usepackage{booktabs}       
\usepackage{amsfonts}       
\usepackage{nicefrac}       
\usepackage{microtype}      
\usepackage{xcolor}         

\usepackage{comment}
\usepackage{multirow}
\usepackage{makecell}
\usepackage[pdftex]{graphicx}
\usepackage{amsmath,amsfonts,amsthm}
\usepackage{mathtools}
\usepackage{adjustbox}
\usepackage{multicol}
\usepackage{multirow}
\usepackage[linesnumbered,ruled]{algorithm2e}
\usepackage{amsmath}
\usepackage{caption}
\usepackage{subcaption}
\usepackage{listings}
\usepackage{wrapfig}
\usepackage{tablefootnote}
\usepackage{tikz}
\usepackage{pifont}

\usepackage{changepage}

\theoremstyle{plain}

\theoremstyle{definition}

\theoremstyle{remark}

\newcommand{\gtiasu}{\texttt{TI-ASU}}
\newcommand{\gtiasus}{\texttt{TI-ASU-S}}
\newcommand{\gtiasumm}{\texttt{TI-ASU-MM}}

\title{TI-ASU: Toward Robust Automatic Speech Understanding through Text-to-speech Imputation Against Missing Speech Modality}

\author{%
  Tiantian Feng$^{1}$, Xuan Shi$^{1}$, Rahul Gupta$^{2}$, Shrikanth S. Narayanan$^{1}$ \\ 
$^1$University of Southern California \quad $^2$Amazon AGI\\
{\small\texttt{tiantiaf@usc.edu}}
{
}
}

\begin{document}

\maketitle

\begin{abstract}
Automatic Speech Understanding (ASU) aims at human-like speech interpretation, providing nuanced intent, emotion, sentiment, and content understanding from speech and language (text) content conveyed in speech. Typically, training a robust ASU model relies heavily on acquiring large-scale, high-quality speech and associated transcriptions. However, it is often challenging to collect or use speech data for training ASU due to concerns such as privacy. To approach this setting of enabling ASU when speech (audio) modality is missing, we propose  \texttt{TI-ASU}, using a pre-trained text-to-speech model to impute the missing speech. We report extensive experiments evaluating \texttt{TI-ASU} on various missing scales, both multi- and single-modality setting, and the use of LLMs. Our findings show that \texttt{TI-ASU} yields substantial benefits to improve ASU in scenarios where even up to 95\% of training speech is missing. Moreover, we show that \texttt{TI-ASU} is adaptive to dropout training, improving model robustness in addressing missing speech during inference.

\end{abstract}

\section{Introduction}
\label{sec:intro}

Speech understanding is fundamental to human communication as the rich information conveyed through speech allows efficient and explicit ways to exchange thoughts, emotions, and ideas. Within the development of conversational AI, automatic speech understanding (ASU) aims to accurately and comprehensively interpret input speech using advanced ML techniques, including bridging potential communication gaps between people from diverse backgrounds and language capabilities.  With the advances in mobile computing, one popular application of ASU has been virtual assistants, such as Amazon Alexa and Apple Siri.

In the deep learning era, the success of data-driven ASU relies heavily on the quality and diversity of the data used for training the models.  To ensure that the ASU model reaches a fair convergence, an ideal corpus is expected to contain an adequate amount of complete audio samples that cover diverse speech attributes, such as emotion, intonation, and speaker demographics \cite{feng2023review}.  However, multiple practical constraints--both technological and human--hinge on the collection of the ASU datasets, such as hardware instability, imbalances in data resources, and the need for privacy protection. To deal with limitations to access speech samples, it is hence necessary to modify the ASU toward robustness to  `missing' speech.

In this work, we focus on investigating the ASU when speech is increasingly missing from the corpus, and in some extreme cases, only the label of the speech is provided during the training.  
Previously, \cite{Zhao2021MissingMI, luo2023multimodal} approached this problem by projecting multi-modal data to a unified feature space from which the missing modality can be reconstructed during inference. Inspired by the recent surge of generative AI, some attempts \cite{he2022synthetic, feng2024can, zhang2023gpt} incorporate synthetic data in the training to boost the model performance in zero-shot setting.  
Moreover, \cite{ueno2021data, fazel2021synthasr, bartelds2023making} have demonstrated that synthetic speech can serve as data augmentation to boost speech recognition.


In this work, we propose \texttt{TI-ASU}, a \textbf{T}ext-to-speech \textbf{I}mputation approach for \textbf{A}utomatic \textbf{S}peech \textbf{U}nderstanding that addresses missing modality challenges in ASU applications. The core idea of \texttt{TI-ASU} is to impute the missing speech modality with TTS models from text transcriptions. Extensive experimental results demonstrate the effectiveness of \texttt{TI-ASU} in various missing scale, single- and multi-modality scenarios.

\begin{figure*}[t]
	\centering
	\includegraphics[width=\linewidth]{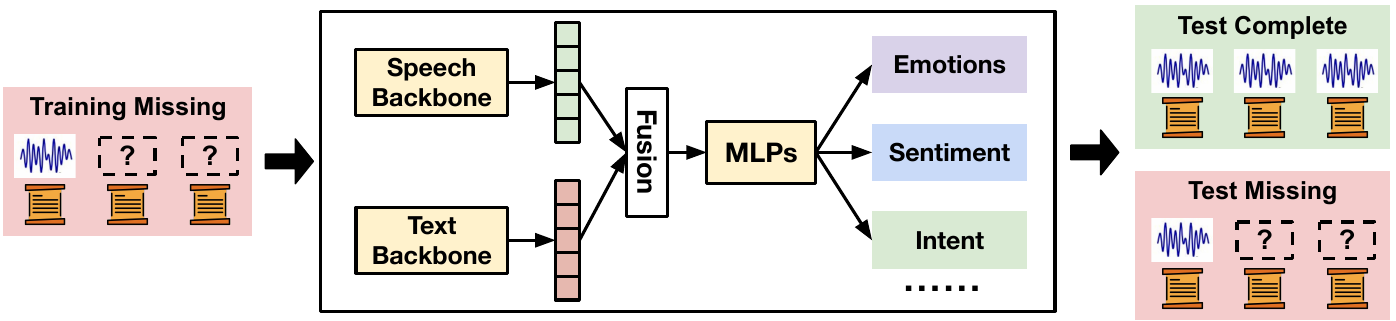}
    \caption{Problem formulation of missing modalities in this work with ASU. The missing speech modality includes cases in training data alone or any data (both training data and testing data).}
    \label{fig:problem_formulation}
    \vspace{-3mm}
\end{figure*}

\vspace{-2mm}
\section{Related Works}

\noindent \textbf{Text-To-Speech (TTS):} TTS aims to generate human-like speech from a given text input. Modern TTS models typically include two modules: 1) an acoustic module that maps intermediate acoustic features from the input text (Tacotrons \cite{wang2017tacotron, shen2018natural}, FastSpeechs \cite{ren2019fastspeech, ren2020fastspeech}, and \textit{et al.}); and, 2) a waveform module that transforms the acoustic feature to the audio (WavNet \cite{oord2016wavenet}, NSF \cite{wang2019neural}, HiFi-GAN \cite{kong2020hifi}, DiffWave \cite{kong2021diffwave}, and \textit{et al.}) With the rise of foundation model in language and speech, SpeechT5 \cite{ao2021speecht5} and VALL-E(X) \cite{wang2023neural, zhang2023speak} learns aligned speech and language representations with self-supervised learning to synthesis speech. 

\vspace{0.5mm}

\noindent \textbf{Missing Modality:} How to leverage a data-driven model with fragmented input? A straightforward method is to remove the incomplete samples \cite{HGMF,Ni2019ModelingHR}.  Instead of removal, another direction is to reconstruct the missing modality from joint representation space \cite{Pham2018FoundIT,Zhao2021MissingMI} or bayesian-meta-learning \cite{Ma2021SMILML}.  With the prevalence of the transformer in multi-model, multiple threads modifications are proposed to extend the transformer for the missing modality input, such as feature-reconstruction \cite{Yuan2021TransformerbasedFR}, tag-based encoding \cite{Zeng2022TagassistedMS}, multi-task modeling \cite{ma2022multimodal}, and prompt learning \cite{lee2023cvpr}. However, existing literature studies this issue from the glass-half-empty perspective, while \texttt{TI-ASU} reframes this problem from the glass-half-full perspective with the assistance of generative AI to boost low-resource speech training.

\vspace{-2mm}

\section{Problem Definition}

\begin{figure}[t]
	\centering
	\includegraphics[width=0.35\linewidth]{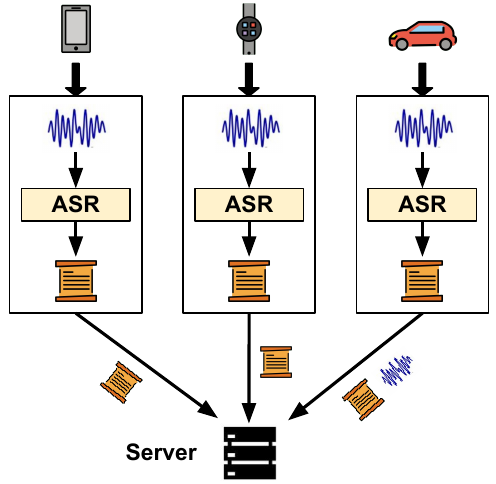}
    \caption{Illustration of edge devices performing ASR services, where text modality is always present.}
    \vspace{-4mm}
    \label{fig:asr_service}
\end{figure}

The present work focuses on studying multi-modal learning for ASU with missing speech modality, as shown in Figure~\ref{fig:problem_formulation} \footnote{The figure in this paper uses images from https://openmoji.org/}. Our motivation to study scenarios with missing speech arises from the fact that speech data carry sensitive information about an individual, such as biometric fingerprints and demographic and health status. In such instances, a common assumption is that the text information may remain accessible by deploying efficient ASR models on the edge, similar to Figure~\ref{fig:asr_service}. However, the speech data is not allowed to egress out of the user devices due to privacy risks, preventing them from being used for training ASU models. Overall, we study two distinct scenarios of missing speech modality: speech missing in training or from any data setting, both in training and testing.


\subsection{Speech-Missing in Training Data}

Our investigation starts with a relatively trivial case where missing speech only occurs in the training data while testing data are with complete speech-text data. Following the notation conventions for missing modalities in \cite{lee2023cvpr}, we define the set of complete training data samples and text-only training samples as $\mathcal{D}_{train}^{C}$ and $\mathcal{D}_{train}^{T}$, respectively. Specifically, we represent $\mathcal{D}_{train}^{C} = \{x_{i}^{T}, x_{i}^{S}, y_{i}\}$ and $\mathcal{D}_{train}^{T} = \{x_{i}^{T}, y_{i}\}$, where $i\in\mathbb{N}$, $T$ and $S$ represent text and speech modalities, respectively. In the end, the training dataset can be expressed as $\mathcal{D}_{train} = \{\mathcal{D}_{train}^{C}, \mathcal{D}_{train}^{T}\}$. Without loss of generality, we introduce the speech modality missing ratio presented in training data, denoted as $p$. Namely, $p=0\%$ indicates the special case that no speech samples are missing. 

\begin{figure*}[t]
	\centering
	\includegraphics[width=\linewidth]{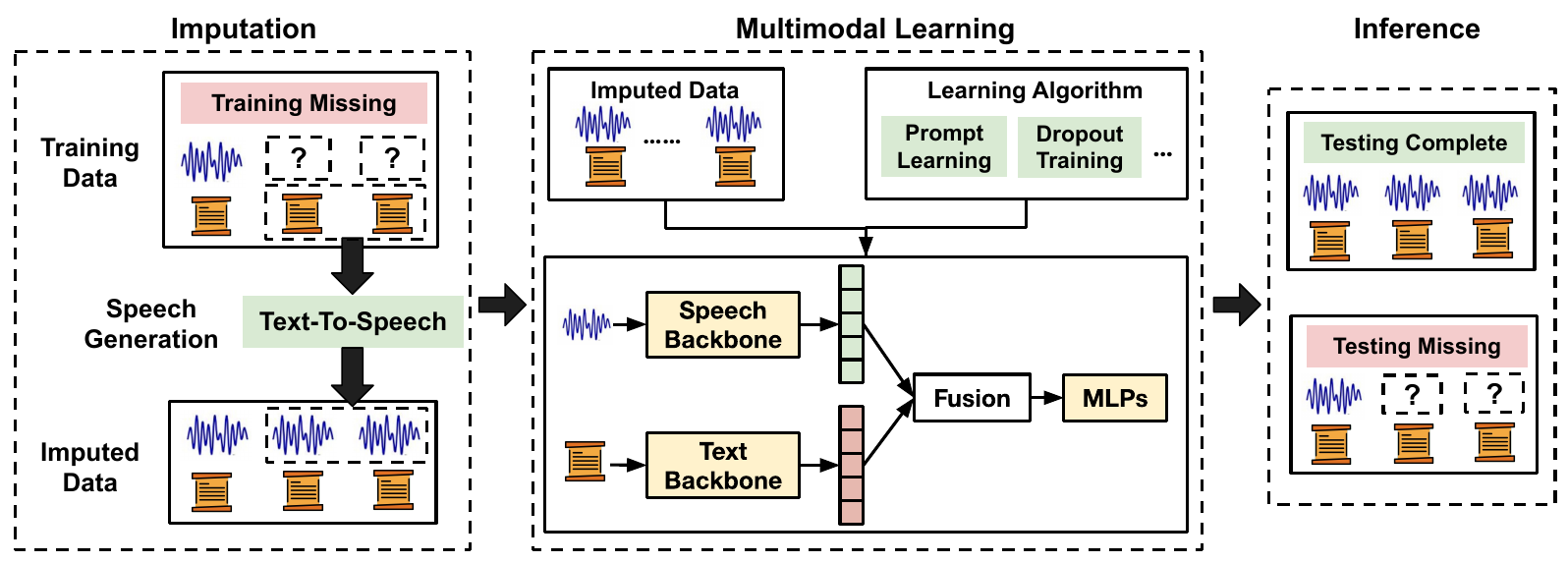}
    \vspace{-2mm}
    \caption{Learning framework of \texttt{TI-ASU}: Imputing missing speech modality with synthetic speech content through text-to-speech transformer models for robust automatic speech understanding.}
    \vspace{-3mm}
    \label{fig:multimodal_leaerning}
\end{figure*}

\subsection{Speech-Missing in Any Data}

Apart from exploring the scenario with speech missing in training data, we further study cases where speech data can be missing in both training and testing data. Previous studies on multimodal learning have highlighted that multimodal models, such as multimodal transformers, are frequently sensitive to modality missing during inference, resulting in substantial performance drops when test samples contain missing modalities. On the other hand, missing modality can be regarded as a robustness challenge in multimodal learning, as the missing modality is a unique case of data perturbation. In summary, we can denote the test dataset as $\mathcal{D}_{test} = \{\mathcal{D}_{test}^{C}, \mathcal{D}_{test}^{T}\}$ following our notation established in the previous subsection. We define the speech modality missing ratio in testing data as $q$, with $q=100\%$ representing the scenario where only text modality is present in the testing set.

\vspace{-1mm}

\section{\texttt{TI-ASU} Framework}

\subsection{Pre-trained Models}

This work experiments with widely adopted pre-trained models as the backbone for modeling ASU, specifically choosing \texttt{WavLM} \cite{chen2022wavlm} and \texttt{RoBERTa} \cite{liu2019roberta} as the speech and text encoders, respectively. The \texttt{WavLM}, a model trained with self-supervised learning, demonstrates promising performance in a wide range of speech-centered tasks. This model is trained with multiple learning objectives, such as frame prediction, speech enhancement, and speech prediction. Our experiments employ the \texttt{WavLM} Base Plus model, consisting of 12 encoding layers with approximately 90M parameters. On the other hand, \texttt{RoBERTa} is a transformer model for text classification based on \texttt{BERT}, benefiting from using a much larger training corpus and modified training objectives than \texttt{BERT}.

\subsection{End-to-End Downstream Modeling}

Our modeling draws inspiration from \cite{feng2023foundation}, highlighting the effectiveness of combining speech and text information for ASU. 

\noindent \textbf{Speech:} Similarly to \cite{pepino2021emotion}, our speech downstream model utilizes learnable weight parameters to combine hidden outputs from all encoder layers. The combined speech output is then fed into two 1D pointwise convolutional layers with a filter size of 256. Finally, global average pooling is applied to the convolutional layer output, resulting in an output vector of size 256. 

\noindent \textbf{Text:} In contrast to speech modeling, we apply a 2-layer bidirectional GRU of a hidden size of 256 to the last hidden output from the \texttt{RoBERTa} model. The output from the GRU layers is averaged to obtain an output embedding of size 256. 

\noindent \textbf{Multimodal Fusion:} In the last stage, we concatenate the text embedding and speech embedding into a multimodal embedding, which is subsequently fed into two fully connected layers for the ASU classification tasks. We want to highlight that speech-only or text-only models respectively feed speech embedding and text embeddings into the classifiers without the fusion.

\subsection{Speech Data Generation}

The core idea behind \texttt{TI-ASU} is to impute the missing speech modality (as shown in Figure~\ref{fig:multimodal_leaerning}) with transformer-based TTS models. To do so, we prompt the TTS models with transcriptions to synthesize speech data. Prior research on visual recognition \cite{Shipard2023DiversityID, he2022synthetic} has demonstrated that diversity is needed in training data generation. For example, researchers propose to include \textbf{multi-domain} knowledge to enrich the diversity of image generation by augmenting the domain "photo" to other domains such as drawing and painting. In contrast to the diversity enhancement approaches in image generation, our proposed \texttt{TI-ASU} enriches the generation diversity by employing multiple TTS models in the generation. This differs from image generation, where diversity is associated with the prompt message. In speech generation, however, the input transcription to the TTS model remains the same. Here, we deploy 3 TTS: Speech-T5 \cite{ao2021speecht5}, MMS-En \cite{pratap2023scaling}, and Vall-E X \cite{zhang2023speak}.

\subsection{Training with Speech Imputation}

The multi-modal learning framework in \texttt{TI-ASU} is presented in Figure~\ref{fig:multimodal_leaerning}. Without loss of generality, we define the generated speech dataset as $\mathcal{D}_{train}^{S'} = \{{x}_{j}^{S'}, y_{j}\}$, where ${x}_{j}^{S'}$ is the generated speech from the text $x_{j}^{T}$ and $S'$ denotes the generated speech. More concretely, given a TTS model $\mathcal{G}$, we can define ${x}_{j}^{S'} = \mathcal{G}(x_{j}^{T})$. We impute each text-only data during each training epoch by randomly selecting a generated speech from the generation set involving three TTS models. Consequently, we obtain a modality-complete dataset $\hat{\mathcal{D}} = \{\mathcal{D}^{C}, \hat{\mathcal{D}}^{T}\}$ with speech data imputation, where $\hat{\mathcal{D}}^{T} = \{x_{i}^{T}, x_{i}^{S'}, y_{i}\}$. Finally, we perform multi-modal training with the imputed dataset $\hat{\mathcal{D}}$. It is worth noting that \texttt{TI-ASU} can integrate with other multi-modal learning algorithms, such as dropout training and prompt learning, as shown in Figure~\ref{fig:multimodal_leaerning}.

\vspace{-2mm}

\section{Datasets}

\begin{table}[t]
    \caption{Summary of dataset statistics used in this work.}
    \footnotesize
    \centering
    \begin{tabular*}{0.52\linewidth}{lccc}
        \toprule
        
        \multirow{1}{*}{\shortstack{\textbf{Datasets}}} & 
        \multirow{1}{*}{\textbf{Speaker \#}} & 
        \multirow{1}{*}{\shortstack{\textbf{Classes}}} &
        \multirow{1}{*}{\shortstack{\textbf{Utterance \#}}}  \\ 

        \midrule
        \textbf{IEMOCAP} & 10 & 4 & 5,531 \\ 
        \textbf{MSP-Improv} & 12 & 4 & 7,798 \\ 
        \midrule
        \textbf{SLURP} & 177 & 46 & 72,277 \\
        \midrule
        \textbf{SLUE} & 140 & 3 & 7,231 \\
        \bottomrule
    \end{tabular*}
    \vspace{-3mm}
    \label{table:datasets}
\end{table}

We utilize four datasets -- IEMOCAP \cite{busso2008iemocap}, MSP-Improv \cite{busso2016msp}, SLURP \cite{bastianelli2020slurp}, and SLUE-Voxceleb \cite{shon2022slue} -- to evaluate \texttt{TI-ASU} across three ASU-related tasks: speech emotion recognition, spoken language understanding, and speech sentiment analysis. Notably, the first two datasets are employed for the identical task.

\section{Experimental Details}

\subsection{Speech Data Generation}

As mentioned previously, we apply 3 TTS experts to generate speech samples: Speech-T5, MMS-En, and Vall-E X. We generate three speech samples for each text-only sample using these TTS models. In particular, in the context of emotion recognition, we generate speech samples by introducing emotion styles associated with the training data when utilizing the Vall-E X model. We would emphasize that Vall-E X derives the emotion styles from datasets distinct from IEMOCAP and MSP-Improv datasets, avoiding introducing in-domain knowledge in the generation process. Most importantly, we choose not to impute missing data in the test set to prevent introducing artifacts in the evaluation.

\subsection{Evaluation and Hyperparameters}

\noindent \textbf{Evaluation:} We use unweighted average recall (UAR) to evaluate emotion recognition, while we utilize the F1 score to evaluate intent and sentiment classifications. We conducted the experiments with 5-fold and 6-fold cross-validation on the IEMOCAP and MSP-Improv datasets, respectively. Moreover, we performed the training with 3 random seeds and reported the average performance on the remaining datasets. For the Slurp dataset, we use standard splits for training, validation, and testing, while for the SLUE-Voxceleb dataset, we split 20\% of the default training data for validation. We report performance on the default validation set, given that the label of the test set is unavailable from the public SLUE-Voxceleb data. 

\noindent \textbf{Hyperparameters:} In all experiments, including fine-tuning for natural and imputed datasets, we set the batch size as 64. Specifically, we set the learning rate at 0.0005 and the maximum training epoch as 30 for training all speech models. We apply a learning rate of 0.0001 and a maximum training epoch of 20 for text and multimodal training, as we observed faster training convergence with text modality. The experiments are conducted on a high-performance computing server with A40 GPUs. We use the checkpoints of each pre-trained model from HuggingFace \cite{wolf2019huggingface}.

\section{Can \textbf{\texttt{TI-ASU}} Improve Data Efficiency with Speech-Missing in Training?}

In this section, we present the results of ASU with speech missing in the training. Specifically, we implement the following approaches for comparisons:

\begin{itemize}[leftmargin=*]
    \item \textbf{Text Training: } Given that a substantial portion of speech is missing from our experiments, one natural baseline is to rely solely on the text for ASU. This baseline is essentially a text-only model.

    \item \textbf{Speech Training: } In addition to the text-only model, another baseline is to train a speech model with the speech data available. Given the missing ratio $p$, we train the model with $100-p\%$ speech.

    \item \textbf{Multimodal Training: } Another extension to speech training with limited speech is adding the paired text. Specifically, given speech missing ratio $p$, we train with $100-p\%$ speech-text data.
    
    \item \textbf{Multi-modal Training with Zero-filling Imputation: } The approaches above use a portion of available data. To utilize all the available data, we impute the missing images by filling them with zeros as in \cite{lee2023cvpr}. For example, we fill $p\%$ speech data with zero in the multi-modal training.

    \item \textbf{\texttt{TI-ASU-S}: } Speech-only training based on \texttt{TI-ASU}.

    \item \textbf{\texttt{TI-ASU-MM}:} Multimodal training based on \texttt{TI-ASU}.
    
\end{itemize}

\begin{table}[t]
    \caption{Comparisons among text, speech, and multi-modal models across different datasets. We compare different models with $p=0\%$, where $p$ indicates speech missing ratio in training.}
    \centering
    \small
    \vspace{0.5mm}
    \begin{tabular*}{0.62\linewidth}{lcccc}
        \toprule
        & 
        \multicolumn{1}{c}{\textbf{Text-Only}} &
        \multicolumn{1}{c}{\textbf{Speech-Only}} & 
        \multicolumn{1}{c}{\textbf{Multi-modal}} \\

        \midrule 
        \textbf{IEMOCAP} & $62.8$  & $69.2$ & $\textbf{71.4}$ \\
        \textbf{MSP-Improv} & $52.7$  & $63.7$ & $\textbf{65.6}$ \\
        \textbf{Slurp} & $78.8$  & $59.2$ & $\textbf{80.3}$ \\
        \textbf{SLUE-Voxceleb} & $\textbf{51.4}$  & $45.1$ & $50.0$ \\

        \bottomrule
    \end{tabular*}
    \vspace{-3mm}
\label{tab:baseline_results}
\end{table}

\subsection{Would text alone be enough for ASU?}

Text modality alone is effective in diverse language-related tasks. Therefore, our investigation begins with whether text data alone achieves competitive ASU performance. To answer this, we compare the text model with speech and multi-modal models trained with complete data, as shown in Table~\ref{tab:baseline_results}. The results demonstrate that multimodal models consistently yield benefits in emotion recognition, while text data exhibits competitive or even better performances compared to multimodal models in other tasks. However, we find that speech models often underperform text and multimodal models. In summary, our experiments highlight, not surprisingly, that speech information can benefit emotion recognition (leveraging the acoustic variation in emotion expression) but does not necessarily contribute to sentiment and intent classifications (which are largely driven by language information).

\begin{table}[t]
    \caption{Comparisons of speech training and TI-ASU-S. Here, TI-ASU-S is trained with pure synthetic data ($p=100\%$), and speech training uses complete real speech for training ($p=0\%$). $p$ indicates the speech missing ratio in training data.}
    \centering
    \small
    \vspace{0.5mm}
    \begin{tabular*}{0.52\linewidth}{lcccc}
        \toprule
        & 
        \multicolumn{1}{c}{\textbf{Speech Training}} & 
        \multicolumn{1}{c}{\textbf{TI-ASU-S}} \\

        & 
        \multicolumn{1}{c}{$\mathbf{p=0\%}$} & 
        \multicolumn{1}{c}{$\mathbf{p=100\%}$} \\

        \midrule 
        \textbf{IEMOCAP} & $69.2$  & $52.7$ \\

        \textbf{MSP-Improv} & $63.7$  & $44.4$  \\

        \textbf{Slurp} & $59.2$  & $55.5$ \\

        \textbf{SLUE-Voxceleb} & $45.1$  & $41.2$\\

        \bottomrule
    \end{tabular*}
    \vspace{-3mm}
\label{tab:zero_shot_speech}
\end{table}

\begin{figure}[t] {
    \centering
    
    \begin{tikzpicture}

        \node[draw=none,fill=none] at (0,0){\includegraphics[width=0.5\linewidth]{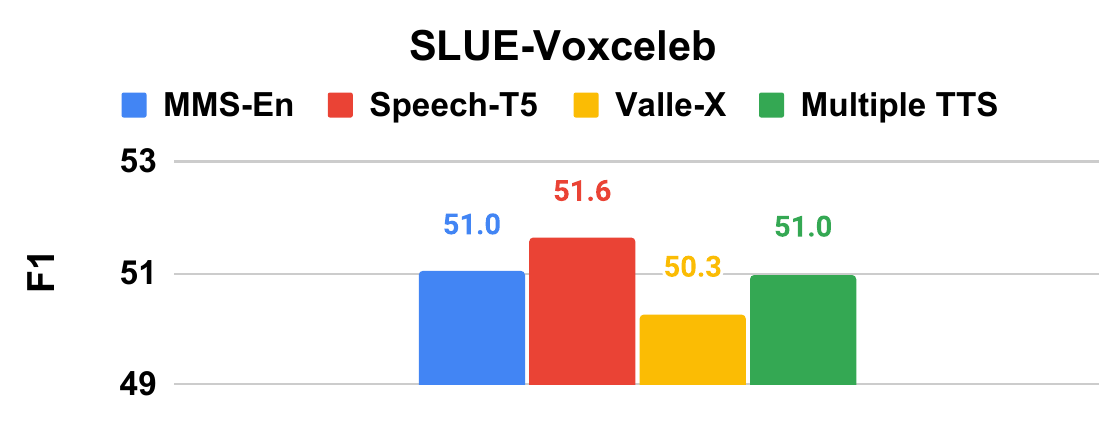}};

        \node[draw=none,fill=none] at (0.5\linewidth,0){\includegraphics[width=0.5\linewidth]{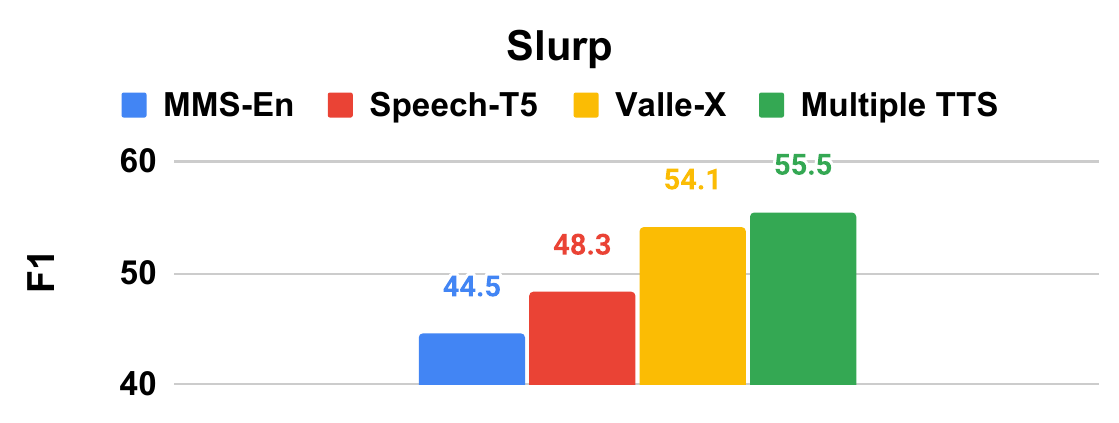}};

        \node[draw=none,fill=none] at (0.5\linewidth,2.7){\includegraphics[width=0.5\linewidth]{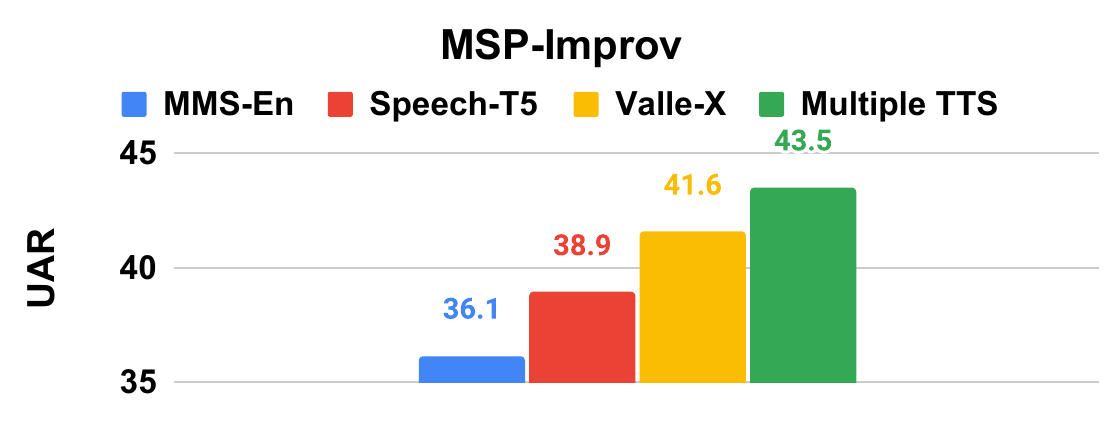}};

        \node[draw=none,fill=none] at (0,2.7){\includegraphics[width=0.5\linewidth]{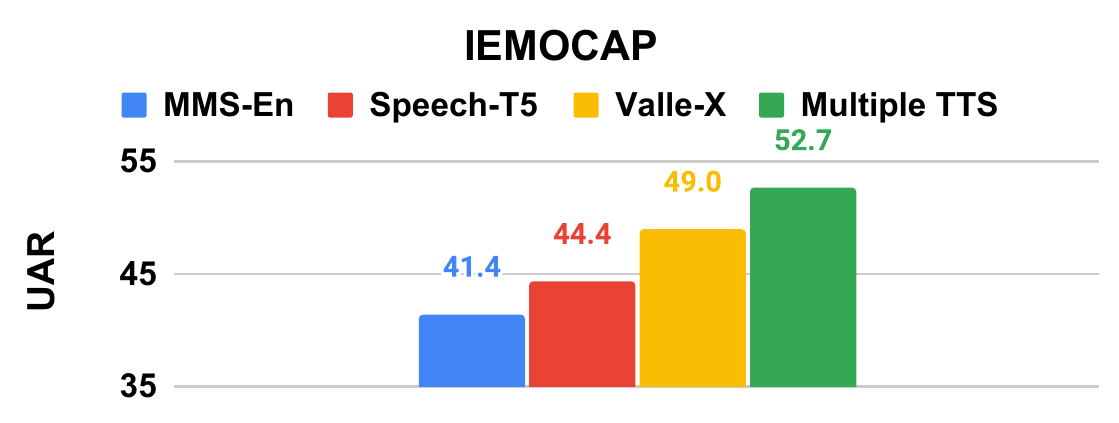}};

    \end{tikzpicture}
    \caption{Comparisons among using single TTS generation and multiple generation in \texttt{TI-ASU}. Here, the training set is entirely based on synthetic speech data.}
    \vspace{-3mm}
    \label{fig:diversity_speech_gen}
} \end{figure}

\subsection{Can \texttt{TI-ASU-S} Provide Competitive Zero-shot Performance Without Real Speech?}

\noindent \textbf{Zero-shot Performance with Synthetic Speech}: While speech data may not consistently benefit multimodal ASU, as shown in Table~\ref{tab:baseline_results}, it is worth investigating whether relying solely on synthetic speech can result in a better-performed speech model. Accordingly, this involves a speech missing ratio $p=100\%$ in training. Table~\ref{tab:zero_shot_speech} compares speech training without missing and \texttt{TI-ASU-S} with $p=100\%$. The results reveal that real speech exhibits substantial advantages in emotion recognition, while synthetic speech alone can yield competitive speech models for intent and sentiment classification. Even though synthetic speech data cannot replace real speech data, our findings suggest their potential as a valuable training resource.

\noindent \textbf{Multiple TTS in Speech Generation}: We further explore the effectiveness of our proposed method in enhancing generation diversity through multiple TTS for ASU training. We compare the \texttt{TI-ASU-S} using individual TTS generation versus our previously mentioned combined approach. The comparisons in Figure~\ref{fig:diversity_speech_gen} suggest that augmenting speech generation by deploying multiple TTS models brings substantial benefits to ASU training, leading to consistent performance increases compared to single TTS generation. Therefore, the remaining results of \texttt{TI-ASU} are based on Multiple TTS generations.

\begin{table}[t]
    \caption{Comparisons of speech models trained using real speech and \gtiasus. Here, we present real speech training and \gtiasus{} with $p=50\%$ and $p=95\%$. $p$ indicates the speech missing ratio in training data.}
    \vspace{1mm}
    \centering
    \footnotesize
    \begin{tabular*}{0.6\linewidth}{lcccc}
        \toprule

        
        & \multicolumn{2}{c}{$\mathbf{p=95\%}$} & \multicolumn{2}{c}{$\mathbf{p=50\%}$} \\
        & \multicolumn{1}{p{0.8cm}}{Speech} & \multicolumn{1}{c}{TI-ASU-S} & \multicolumn{1}{p{0.8cm}}{Speech} & \multicolumn{1}{c}{TI-ASU-S} \\

        \cmidrule(lr){1-1}
        \cmidrule(lr){2-3}
        \cmidrule(lr){4-5}
        \textbf{IEMOCAP} & $52.5$ & $\mathbf{54.8}$ & $65.7$ & $\mathbf{65.8}$ \\

        \textbf{MSP-Improv} & $44.4$ & $\mathbf{48.3}$ & $\mathbf{61.5}$ & $60.5$ \\

        \textbf{Slurp}  & $<5\%$ & $\mathbf{56.0}$ & $50.4$ & $\mathbf{58.0}$ \\

        \textbf{SLUE-Voxceleb} & $34.4$ & $\mathbf{41.2}$ & $40.4$ & $\mathbf{45.7}$ \\

        \bottomrule
    \end{tabular*}
    \vspace{-3mm}
\label{tab:low_resource_speech}
\end{table}

\begin{table}[t]
    \caption{Comparing \texttt{TI-ASU-MM} with models trained using complete text and multimodal data. Here, \texttt{TI-ASU-MM} is with $p=100\%$, meaning no real speech is used in training but synthetic speech.}
    \vspace{1mm}
    \centering
    \footnotesize
    \begin{tabular*}{0.55\linewidth}{lcccc}
        \toprule
        & \multirow{2}{*}{\textbf{Text}} & 
        \multirow{2}{*}{\textbf{Multimodal}} & 
        \multicolumn{1}{c}{\textbf{TI-ASU-MM}} \\

        & & & 
        \multicolumn{1}{c}{$\mathbf{p=100\%}$} \\

        \midrule 
        \textbf{IEMOCAP} & $62.8$  & $71.4$ & $67.1$ \\

        \textbf{MSP-Improv} & $52.7$  & $65.6$ & $58.0$ \\

        \textbf{Slurp} & $78.8$  & $80.3$ & $79.4$\\

        \textbf{SLUE} & $51.4$  & $50.0$ & $51.0$\\

        \bottomrule
    \end{tabular*}
    \vspace{-1mm}
\label{tab:zero_shot_mm}
\end{table}

\subsection{Can \texttt{TI-ASU-S} Improve Model Performance with  Limited Real Speech?}

Table~\ref{tab:zero_shot_speech} shows the importance of real speech data in training ASU, leading us to study training ASU models with limited available speech data. $p$ indicates the speech-missing ratio in training.

\noindent \textbf{\texttt{TI-ASU-S} with $\mathbf{p=95\%}$} (extreme missing condition):  To begin with, we study severe speech-missing conditions, including cases where the speech-missing ratio is $p=95\%$. Table~\ref{tab:low_resource_speech} compares the performance between speech training and \texttt{TI-ASU-S} at $p=95\%$. The results indicate that our proposed \texttt{TI-ASU-S} can consistently outperform real-speech training, where $95\%$ of speech is missing. We identify that this performance increase is substantial in intent classification, where the speech training fails to converge.

\noindent \textbf{\texttt{TI-ASU-S} with $\mathbf{p=50\%}$} (moderate missing condition): In addition to cases where a significant portion of the speech data is missing, we extend our investigation to less severe scenarios where only half of the speech is unavailable. The results in Table~\ref{tab:low_resource_speech} demonstrate that even with more real speech, \texttt{TI-ASU-S} can yield better performances compared to real speech training in most datasets. Similar to results in cases where $p=95\%$, we identify that \texttt{TI-ASU-S} provides notable performance improvements in intent classification. However, the performance benefit is smaller at $p=50\%$ compared to $p=95\%$.

\begin{table}[t]
    \caption{Comparing \texttt{TI-ASU-MM} with multimodal learning using available speech-text pairs and zero-filling imputation on missing speech. Here, we experiment with the condition where $p=95\%$. $p$ indicates the speech missing ratio in training data.}
    \centering
    \vspace{1mm}
    \footnotesize
    \begin{tabular*}{0.56\linewidth}{lcccc}
        \toprule
        & 
        \multirow{2}{*}{\textbf{MM}} & 
        \multicolumn{1}{c}{\textbf{MM Zero Filling}} & 
        \multirow{2}{*}{\textbf{TI-ASU-MM}} \\

        & & \textbf{Imputation} & \\

        \midrule 
        \textbf{IEMOCAP} & $52.8$  & $63.9$ & $\mathbf{67.1}$ \\
        \textbf{MSP-Improv} & $32.8$  & $53.2$ & $\mathbf{57.9}$ \\
        \textbf{Slurp} & $45.9$  & $78.7$ & $\mathbf{80.4}$ \\
        \textbf{SLUE} & $42.5$  & $\mathbf{51.4}$ & $50.6$\\

        \bottomrule
    \end{tabular*}
    \vspace{-3mm}
\label{tab:low_resource_mm}
\end{table}

\begin{figure}[t] {
    \centering
    
    \begin{tikzpicture}

        \node[draw=none,fill=none] at (0.5\linewidth,0){\includegraphics[width=0.5\linewidth]{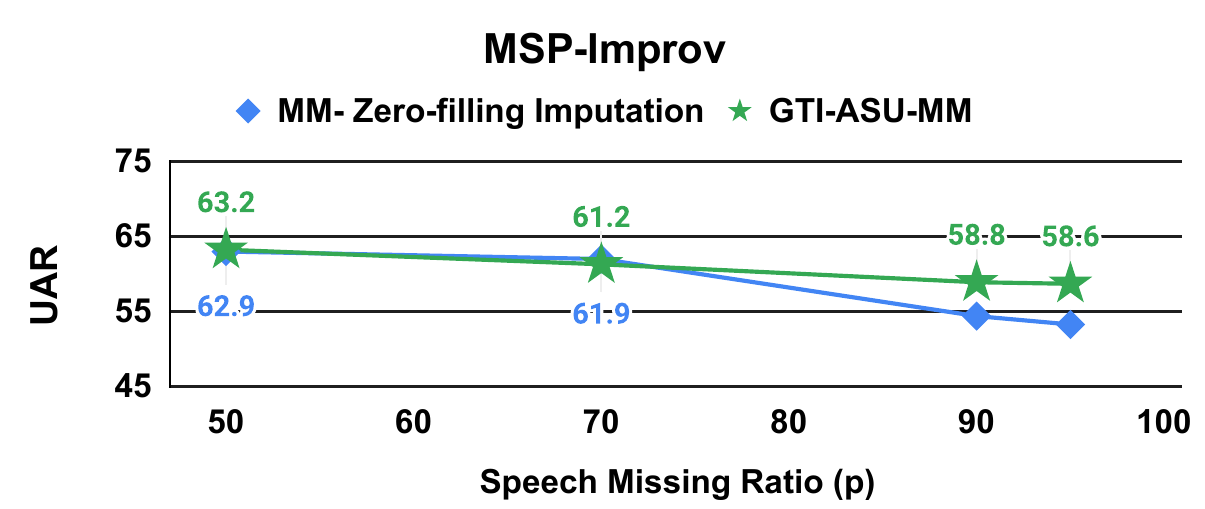}};

        \node[draw=none,fill=none] at (0,0){\includegraphics[width=0.5\linewidth]{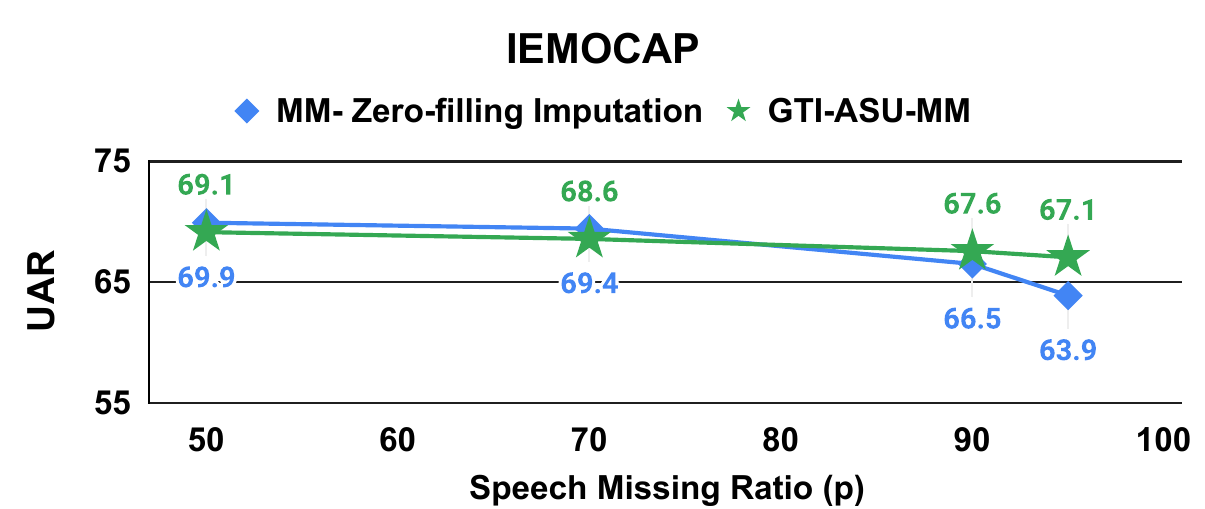}};

    \end{tikzpicture}
    \caption{Comparisons of multimodal training between \texttt{GTI-ASU} and zero-filling imputation on missing speech. Speech missing ratio in training data $p\in\{50\%, 70\%, 90\%, 95\%\}$}
    \label{fig:mm_gti_asu_p}
} \end{figure}

\subsection{Can \texttt{TI-ASU-MM} Provide Competitive Multimodal Models with No Real Speech?}

Apart from assessing \gtiasu{} in assisting speech training for ASU, we further investigate multimodal learning, incorporating text modality into ASU training. Similar to \texttt{TI-ASU-S} experiments, we begin with \gtiasumm where speech missing ratio $p=100\%$, indicating that no real speech is available in the training. Table~\ref{tab:zero_shot_mm} compares \gtiasumm, involving pure synthetic speech with text, against text-only and multimodal training. The latter two are with complete data in training. The results indicate that \gtiasumm{} can improve emotion recognition compared to text-only training, leading to 5-6\% performance increase. However, \gtiasumm{} underperforms multimodal learning in emotion recognition. On the other hand, we find that \gtiasumm{} does not improve intent or sentiment classification, where the text-only model provides the best performances.

\subsection{Can \texttt{TI-ASU-MM} Improve Multimodal Performance with Limited Real Speech?}

\noindent \textbf{\texttt{TI-ASU-MM} with $\mathbf{p=95\%}$} (extreme missing condition): Similar to the \gtiasu{} experiments where we compared \gtiasu{} with speech training of limited speech data, we compare \gtiasumm{} with multimodal training with data that both modalities are present. In addition, we compare \gtiasumm{} with multimodal training using zero-filling imputation on missing speech samples as shown in Table~\ref{tab:low_resource_mm}. Similar to findings from speech training, we observe that \gtiasumm{} benefits emotion recognition in multimodal learning the most. In addition, this advantage is also exhibited in intent classification but not in sentiment classification. Overall, the results support that \gtiasu{} enhances ASU performance in multimodal learning by imputing missing speech with TTS models.

\vspace{0.5mm}
\noindent \textbf{\texttt{TI-ASU-MM} with Lower $\mathbf{p}$}: Moreover, we assess \gtiasumm{} where limited real speech is accessible. Given our findings suggesting marginal or no benefits of \gtiasu{} in intent and sentiment classification, we focus this analysis on emotion recognition. We plot the comparisons between \gtiasumm{} and multimodal learning with zero-filling imputation on IEMOCAP and MSP-Improv, as shown in Figure~\ref{fig:mm_gti_asu_p}. The plot highlights that \gtiasumm{} is more effective in enhancing model performance when the speech missing ratio is significant (e.g., $p>70\%$), and multimodal learning with zero-filling imputation delivers comparable performances to \gtiasumm{} when $50\%$ of speech is available.

\section{Can \textbf{\texttt{TI-ASU}} Enhance Robustness with Speech-Missing in Train and Test?}

This section extends our previous experiment from the speech missing in training data to any data. Dropout training \cite{ma2022multimodal} has been widely applied to enhance the robustness of the multimodal models against missing modalities during inference. In the dropout training, a random portion of the selected modality is replaced with zeros for each batch. In this paper, we perform dropout on speech modality. Specifically, we propose \textbf{\texttt{TI-ASU Dropout}}, combing the dropout training with \texttt{TI-ASU}, where we randomly fill the data in the imputed dataset $\hat{\mathcal{D}} = \{\mathcal{D}^{C}, \hat{\mathcal{D}}^{T}\}$ with zero. Moreover, our baseline is \textbf{MM-Dropout}, multimodal dropout training with complete data, which is the baseline in \cite{lee2023cvpr, ma2022multimodal}. The dropout training uses the dropout rate equal to the test speech missing ratio $q$. We evaluate the model with the missing ratio $q=\{50\%, 70\%, 90\%\}$.

\begin{table}[t]
    \caption{Comparing MM-Dropout (Multimodal-Dropout) with \texttt{TI-ASU Dropout}. $p$ and $q$ are training and testing speech missing ratios, respectively.}
    \centering
    \vspace{1mm}
    \footnotesize
    \begin{tabular*}{0.55\linewidth}{cccccc}
        \toprule
        \multirow{2}{*}{\textbf{Dataset}} & 
        \multirow{2}{*}{\textbf{Dropout}} &
        \multirow{2}{*}{$\mathbf{p}$} &
        \multicolumn{3}{c}{\textbf{Test Missing Ratio ($q$)}} \\
        
        & & &
        \multicolumn{1}{c}{\textbf{50}} &
        \multicolumn{1}{c}{\textbf{70}} &
        \multicolumn{1}{c}{\textbf{90}} \\

        \cmidrule(lr){1-6}
        \multirow{3}{*}{\textbf{IEMOCAP}} & \textbf{MM} & $0$ & $\mathbf{66.6}$ & $64.5$ & $63.8$ \\

        \cline{2-3} & \multirow{2}{*}{\textbf{TI-ASU}} & $50$ & $66.4$ & $\mathbf{64.6}$ & $\mathbf{64.1}$  \rule{0pt}{2.25ex} \\
        & & $95$ & $63.9$ & $63.2$ & $63.3$ \\

        \cmidrule(lr){1-6}
        
        \multirow{3}{*}{\textbf{Slurp}} & \textbf{MM} & $0$ & $79.0$ & $78.3$ & $78.6$ \\

        \cline{2-3}
        & \multirow{2}{*}{\textbf{TI-ASU}} & $50$ & $78.4$ & $78.1$ & $78.2$ \rule{0pt}{2.25ex} \\
        & & $95$ & $\mathbf{79.4}$ & $\mathbf{78.8}$ & $\mathbf{79.0}$ \\

        \cmidrule(lr){1-6}
        
        \multirow{3}{*}{\textbf{SLUE}} & \textbf{MM} & $0$ & $51.0$ & $49.3$ & $50.4$ \\

        \cline{2-3}
        & \multirow{2}{*}{\textbf{TI-ASU}} & $50$ & $\mathbf{51.4}$ & $50.6$ & $\mathbf{51.3}$ \rule{0pt}{2.25ex} \\
        & & $95$ & $51.4$ & $\mathbf{51.1}$ & $51.1$ \\

        \bottomrule
    \end{tabular*}
    \label{tab:dropout_ti_asu}
    \vspace{-3mm}
\end{table}

\begin{table}[t]
    \caption{Comparisons of \texttt{TI-ASU} using original transcript and LLM-assisted transcript augmentation. Here, we present both \texttt{TI-ASU-S} and \texttt{TI-ASU-MM} with $p=95\%$. $p$ indicates the speech missing ratio in training data.}
    \centering
    \vspace{1mm}
    \scriptsize
    \begin{tabular*}{0.9\linewidth}{lcccccc}
        \toprule
        & \multirow{2}{*}{Speech} & \multirow{2}{*}{TI-ASU-S} & \multirow{2}{*}{TI-ASU-S (LLM)} 
        & \multicolumn{1}{c}{MM Zero-Filling} & \multirow{2}{*}{{TI-ASU-MM}} & \multirow{2}{*}{{TI-ASU-MM} (LLM)} \\

        & & & & \multicolumn{1}{c}{Imputation} & & \\
        
        \cmidrule(lr){1-1}
        \cmidrule(lr){2-4}
        \cmidrule(lr){5-7}
        \textbf{IEMOCAP} & $52.5$ & $\mathbf{54.8}$ & $54.5$ & $63.9$ & $\mathbf{67.1}$ & $65.4$ \\

        \textbf{Slurp}  & $<5\%$ & $\mathbf{56.0}$ & $50.4$ & $78.7$ & $\mathbf{80.4}$ & $76.2$\\

        \textbf{SLUE} & $34.4$ & $41.2$ & $\mathbf{41.7}$ & ${51.4}$ & $50.6$ & $\mathbf{53.0}$ \\

        \bottomrule
    \end{tabular*}
    \vspace{-3mm}
\label{tab:low_resource_speech_llm}
\end{table}

\vspace{0.5mm}

\noindent \textbf{Can \texttt{TI-ASU Dropout} outperform MM-Dropout when $p=95\%$ (extreme missing condition)?} Here, we train \texttt{TI-ASU} with $p=95\%$, meaning that only 5\% of real speech is available. The comparisons between MM-Dropout and \texttt{TI-ASU Dropout} are listed in Table~\ref{tab:dropout_ti_asu}. The results show that \texttt{TI-ASU} with dropout training is more robust than the multi-modal dropout training in intent and sentiment classifications. However, we observe this difference is marginal. In contrast, MM dropout outperforms \texttt{TI-ASU Dropout} in emotion recognition when $q$ is low. 

\noindent \textbf{Can \texttt{TI-ASU Dropout} outperform MM-Dropout when $p=50\%$ (moderate missing condition)?} We further investigate \texttt{TI-ASU Dropout} with more real data in training. The comparison in Table~\ref{tab:dropout_ti_asu} shows that \texttt{TI-ASU Dropout} at $p=50\%$ yields competitive emotion recognition performance compared to MM-dropout at different $q$. In addition, we find marginal differences between \texttt{TI-ASU Dropout} and multimodal dropout in the remaining tasks. Overall, these results suggest that, at a lower $p$, \texttt{TI-ASU Dropout} enhances model robustness against test speech missing.


\vspace{0.5mm}

\section{Can LLM-Assisted Speech Generation Apply to \textbf{\texttt{TI-ASU}}?}

\subsection{Generation Approach}
In the previous section, we investigated speech imputation using speech generated from transcriptions. We argue that the diversity of this generation process is limited due to the constrained language content from the original speech. To augment existing speech generation, we propose to augment the transcriptions using the recently released LLM, LLaMa2-70B \cite{touvron2023llama}. In particular, we adopt the following prompting message to augment the text transcriptions: 

\noindent \texttt{Don't repeat my instructions. 
Rephrase the following sentence: 
TRANSCRIPTIONS}

Specifically, for the text-only training set $\mathcal{D}_{train}^{T} = \{x_{i}^{T}, y_{i}\}$, LLM model $\mathcal{F}$, and prompt function $prompt(\cdot)$, we create augmented text set $\mathcal{D}_{train}^{T_{aug}} = \{x_{i}^{T_{aug}}, y_{i}\}$, where $x_{i}^{T_{aug}}=\mathcal{F}(prompt(x_{i}^{T}))$. Moreover, we perform a similar approach to generate speech set $\hat{\mathcal{D}}^{T_{aug}} = \{x_{i}^{T_{aug}}, x_{i}^{S'}, y_{i}\}$. Finally, we combine $\hat{\mathcal{D}}^{T_{aug}}$ and $\hat{\mathcal{D}}^{T}$ to form the imputed multimodal set $\hat{\mathcal{D}}^{T'}$ with speech missing. During training, we randomly select a pair of text and speech from $\hat{\mathcal{D}}^{T'}$ as the data for training ASU. In this way, we enrich the generation diversity by employing multiple TTS models and language content enhancement using LLM.

\subsection{Generation Examples}
Here, we provide generation examples from IEMOCAP, Slurp, and SLUE-Voxceleb datasets. 

\vspace{1mm}
\noindent \textbf{Dataset - IEMOCAP}

\noindent \texttt{Raw: Out there somewhere there is this huge mass of silverfish headed this way}

\noindent \texttt{Augmented: A large school of shimmering silver fish is making its way towards us.}

\noindent \textbf{Dataset - Slurp}

\noindent \texttt{Raw: i did not want you to send that text yet wait until i say send}

\noindent \texttt{Augmented: I wanted you to hold off on sending the text message until I gave you the green light, rather than sending it right away.}

\noindent \textbf{\texttt{SLUE-Voxceleb}}

\noindent \texttt{Raw: well it's hard to get the the right the proper finance money to}

\noindent \texttt{Augmented: Securing adequate financial resources can be a challenge.}

\subsection{Performance of LLM-assisted TI-ASU}

We compare the LLM-assisted TI-ASU with TI-ASU that relies on the original transcript. We conduct the experiments using speech and multimodal training with $p=95\%$. Particularly, we perform the LLM augmentation on IEMOCAP, Slurp, and SLUE-Voxceleb datasets, as demonstrated in Table~\ref{tab:low_resource_speech_llm}. The results indicate that LLM augmentation does not always improve the fine-tuning performances, yielding decreased performance compared to \texttt{TI-ASU} in IEMOCAP and Slurp datasets. We notice this performance drop exists in both speech and multimodal training. However, we observe that the LLM-assisted augmentation can benefit sentiment classification, improving performance on SLUE datasets in both speech and multimodal training. Overall, the findings suggest that LLM holds promise for augmenting speech samples. However, there is a need to enhance the quality of the rephrased transcription utilized for speech generation. This can be achieved through designing more effective prompt strategies or employing more advanced models such as ChatGPT.

\section{Conclusion}
In this work, we proposed \texttt{TI-ASU}, a TTS imputation approach, for multi-modal ASU learning to address the challenges caused by missing speech modality. Our experiments demonstrate that \texttt{TI-ASU} provides robust multi-modal solutions against severe missing speech modality settings in training data or testing data. Crucially, increasing the diversity through multiple TTS speech generation enhances \texttt{TI-ASU} performance. 

\noindent \textbf{Limitations and Future Works:} While \texttt{TI-ASU} is effective in imputing speech data using raw transcript for ASU training, it encounters challenges in combining LLM-assisted augmentations, implying the potential issues in text generation. For example, we identify that LLM frequently extends the spoken language content, making it unlikely to be used in daily communication. Moreover, enriching speaker diversity in speech generation remains a challenge. Our future work plans to explore more advanced LLMs to improve the quality of the text augmentation. Moreover, we plan to include human inspection to evaluate generation quality.

\bibliographystyle{plain}
\bibliography{mybib}



\end{document}